\documentclass[11pt]{article}
\usepackage{epsfig}
\usepackage{a4,isolatin1}
\usepackage{amsmath,amsfonts,latexsym, amssymb}

\newtheorem{satz}{Theorem}[section]
\newtheorem{defi}[satz]{Definition}

\newtheorem{bem}[satz]{Remark}
\newtheorem{lemma}[satz]{Lemma}
\newtheorem{koro}[satz]{Corollary}

\newtheorem{conclusion}[satz]{Conclusion}
\newtheorem{ob}[satz]{Observation}

\newtheorem{propo}[satz]{Proposition}

\newcommand{\mcal}{\mathcal}

\newcommand{\tit}{\textit}

\newcommand{\N}{\mathbb{N}}
\newcommand{\R}{\mathbb{R}}
\newcommand{\Z}{\mathbb{Z}}

\newcommand{\bewende}{$ \hfill \Box $}

\begin{document}
\thispagestyle{empty}
\begin{center}
\vspace*{1.0cm}

{\LARGE{\bf The Continuum Limit of Discrete Geometries}} 

\vskip 1.5cm

{\large {\bf Manfred Requardt }} 

\vskip 0.5 cm 

Institut f\"ur Theoretische Physik \\ 
Universit\"at G\"ottingen \\ 
Friedrich-Hund-Platz 1 \\ 
37077 G\"ottingen \quad Germany\\
(E-mail: requardt@theorie.physik.uni-goettingen.de)

\end{center}

\vspace{1 cm}

\begin{abstract}
In various areas of modern physics and in particular in quantum
gravity or foundational space-time physics it is of great importance
to be in the possession of a systematic procedure by which a
macroscopic or continuum limit can be constructed from a more
primordial and basically discrete underlying substratum, which may
behave in a quite erratic and irregular way. We develop such a
framework within the category of general metric spaces by combining
recent work of our own and ingeneous ideas of Gromov et al, developed
in pure mathematics. A central role is played by two core
concepts. For one, the notion of intrinsic scaling dimension of a
(discrete) space or, in mathematical terms, the growth degree of a
metric space at infinity, on the other hand, the concept of a metrical
distance between general metric spaces and an appropriate scaling
limit (called by us a geometric renormalisation group) performed in
this metric space of spaces. In doing this we prove a variety of
physically interesting results about the nature of this limit process,
properties of the limit space as e.g. what preconditions qualify it as
a smooth classical space-time and, in particular, its
dimension.\\[0.3cm]
Keywords: Gromov-Hausdorff-Distance, Discrete Metric Spaces, Continuum
Limit, Geometric Growth at Infinity, Dimension
\end{abstract} \newpage
\setcounter{page}{1}

\section{Introduction}
In this note we want to report on interesting parallel structures in,
on the one hand, a sub-field of current research in quantum gravity or
foundational (quantum) space-time physics and, on the other hand, in
certain fields of modern mathematics we became aware of only recently
(cf.  the two beautiful essays by M.Berger, \cite{Berger}, about the
work of Gromov in geometric group theory and related fields).

In various areas of modern physics and, in particular, in quantum
gravity it is extremely important to develop effective (technical)
methods which allow one to perform a suitable continuum limit,
starting from discrete and frequently quite irregular structures like
e.g. (fluctuating) large and densely netted networks or dynamic
graphs.

The underlying physical idea is, that (quantum) space-time on the
Planck scale is presumably a very erratic and wildly fluctuating
structure which does not resemble very much anything akin to the
macroscopic space-time continuum we are accustomed to. One might
therefore entertain the idea to model this primordial substratum as a
basically discrete dynamic structure of relatively elementary
constituents and their interactions and try to reconstruct our
continuum concepts by way of performing an appropriate continuum
limit, shedding, by the same token, some light on the deep question
which discrete concepts are the appropriate counterparts of their
continuum cousins.

We note that this problem is virulent in all present approaches to
quantum gravity in one form or the other, the most wide spread being
string theory and loop quantum gravity (see for example
\cite{Markopoulou},\cite{Oeckl} or \cite{Seriu}). There exist however
other, less well-known, but nevertheless promising frameworks. While
taking an appropriate continuum limit is a desideratum in all the
different approaches, the strategies employed are, on the other hand,
too different concerning concepts and technical details so that we
prefer to concentrate in the following on a particular approach we
developed in the recent past and which does not need too many extra
assumptions. In contrast to other frameworks which are implicitly or
explicitly mainly inspired by ideas, occurring in combinatorial
topology and related fields (simplicial complexes to give an example),
we rather view primordial space-time as a large dynamic array of
interacting elementary degrees of freedom which have the propensity to
generate, as an emergent phenomenon, a macroscopic smooth space-time
on a coarser level of resolution. It has the further advantage that
the mathematical framework, we are going to develop in the following,
becomes quite transparent and coherent. Furthermore, it is
sufficiently general and flexible so as to be applicable, after slight
modifications, to related problems in a wide variety of other
contexts.

This change of working philosophy implies that in the following we
will mainly work in the metric category of general spaces and not so
much in the differential geometric category. We think these metrical
aspects are of a more primordial character in physics and are standing
between the more pristine topological aspects of a theory and the more
advanced analytical properties which are perhaps only very useful
(over)idealisations (see also \cite{Isham}). Our general aim will be
the study of large discrete networks assumed to model space-time on a
primordial level, where by discrete we do not necessarily mean
something like lattices or other countable structures but rather the
absence of the usual continuum concepts. We note in particular that it
is important to avoid any idea of an ambient space in which our
networks are assumed to be embedded.  In studying model systems it is
of course frequently advantageous to restrict the analysis (mostly for
technical convenience) to countable structures.  We do not want to
comment on the pros and cons of our general working philosophy and
refer instead to some of our recent papers (see for example
\cite{Req1},\cite{Req2},\cite{Req3}) where more references can be
found. We emphasize however that the framework we present in the
following is quite self-contained and does not need many prerequisites
coming from elsewhere. While being based on a different set of
fundamental principles, the work of Borchers and Sen shares some of
the aims, pursued by us, for example, reconstructing continuum
concepts of space-time from more fundamental notions (see
\cite{Borchers1},\cite{Borchers2}).

Sufficiently interesting and complex examples of discrete geometric
structures are large or even infinite graphs. They represent a rich
class of models we employed in the description of aspects of the
underlying geometric substratum of our networks in the above cited
papers and they will also serve as important model geometries in our
following analysis.

Just for the record we give some definitions. While in some of our
papers (for example in \cite{Req4}) orientation and direction of edges
played a role for doing discrete (functional) analysis, this is, for
the time being, not relevant for the following more geometric
analysis. Furthermore we deal only with what some graph theorists call
\tit{simple graphs}.
\begin{defi}A countable labelled (unoriented) graph, $G=(V,E)$,
  consists of a countable set of vertices (or nodes), $x_i\in V$, and
  a countable set of edges (or links), $e_{ij}=(x_i,x_j)\in E\subset
  V\times V$ so that $e_{ij}$ and $e_{ji}$ are identified, put
  differently, the relation $E$ is assumed to be symmetric.
\end{defi}
Remark: It is important to note that in general we consider abstract
graphs, i.e. not assumed to be embedded in an ambient topological
space. Both vertices and edges can be considered as representatives of
concepts or properties which, in principle, have nothing to do with
points and lines in some ambient geometric environment. In our working
philosophy the edges frequently represent elementary interactions
between certain
degrees of freedom or information channels.\\[0.2cm]
For a general orientation see for example \cite{Bollo1}. At the moment
we introduce only very few concepts from graph theory, making further
comments later when needed.
\begin{defi}A vertex, $x$, has vertex degree $v(x)\in \N_0$ or
  $=\infty$ if it is incident with $v(x)$ edges, put differently, the
  number of unoriented pairs $(x,y)$, in which $x$ occurs. The degree
  function $v(x)$ is called locally bounded if $v(x)\in \N_0$. It is
  called globally bounded if $v(x)\leq v<\infty$ on $V$. The graph is
  called regular if $v(x)$ is constant on $V$.
\end{defi}
Remark: In the functional analysis or operator theory on graphs,
graphs with locally or globally bounded vertex degree represent
important subclasses.\vspace{0.2cm}

Whereas this is not really necessary we deal in the following for
convenience with \tit{connected} graphs.
\begin{defi}A graph is called connected if each pair of nodes, $x,y$ can be
  connected by a finite edge sequence, $\gamma$, starting from $x$ and
  ending in $y$. A path is an edge sequence without repetition of
  vertices with the possible exception of initial and terminal
  vertex. The number of edges occurring in the path is called the
  (combinatorial) length, $l(\gamma)$, of the path. 
\end{defi}
\begin{ob}As these numbers are integer, there exists always a
  path of minimal length, called a geodesic path. This defines a
  natural distance concept on graphs (path metric).
\begin{equation}d(x,y):=\min_{\gamma}\{l(\gamma),\gamma\;\text{connects}\;x\;\text{with}\;y\}
\end{equation}
has the properties of a metric, i.e.
\begin{equation}d(x,x)=0\;,\;d(x,y)=0\rightarrow
  x=y\;,\,d(x,y)=d(y,x)\,,\,d(x,z)\leq d(x,y)+d(y,z)
\end{equation}
the last relation being the crucial one.
\end{ob}
The proof is more or less obvious.\\[0.2cm]
Remark: With this metric $G$ becomes a complete, discrete metric
space. We note however that one can introduce many other types of
metrics on graphs. A simple choice is the transition from $d$ to
$\lambda\cdot d\,,\,\lambda\in \R$ (a metric playing a certain role in
the following). More intricate types of metrics have for example been
studied in \cite{Req5}.\vspace{0.2cm}

The notion of dimension is an important characteristic in continuum
mathematics and physics. In our context it is important to define
related concepts on discrete disordered structures which go over in
corresponding continuum concepts when performing some continuum limit.
On the other hand, in physics it is crucial that these concepts encode
certain relevant geometric characteristics of the physical systems
under discussion. We formulated such a concept in \cite{Req6} and
showed its usefulness. In \cite{Req2} we used it among other concepts,
to characterize the nature of what we called a \tit{geometric
  renormalisation process}, i.e. a coarse-graining and rescaling
process which, hopefully, allows us to construct a continuum limit,
the supposed \tit{fixed point} of the process, from a discrete
underlying substratum. The idea has some vague similarities to the
\tit{real-space renormalisation group} of statistical mechanics, but
it represents, as far as we can see, a technically much more
complicated enterprise. 

In the following two sections we will mainly study these dimensional
concepts and relate them to important concepts being used in
\tit{geometric group theory} and related fields in pure mathematics.
With applications in physics in our mind, we will mainly be interested
in situations where the underlying primordial structure, i.e. in our
case the large irregular graphs or networks, has a definite
generalized dimension (which, as we will see, is already a particular
situation). This holds the more so if this dimension happens to be an
integer. Our aim is, among other things, to provide criteria on this
fundamental scale under which these properties do occur. In the
remaining sections we will show that what we defined as a geometric
renormalisation process has a strong and surprising relation to a
relatively recent and intensively studied topic of pure mathematics
which is connected with the name of M.Gromov, i.e. the study of the
asymptotic limits of general metric spaces, particular subclasses
being finitely generated groups and their Cayley graphs and fractal
geometries. The combination of these tools and concepts allow us to
prove some interesting results about existence and properties of
continuum limits of discrete spaces. To sum it up, on the more
technical side of the matter, the two topics we are going to study are
first, generalized dimension as a characteristic of classes of general
metric spaces and second, distances between metric spaces and possible
limits of sequences of such spaces.
\section{Dimension and Growth on Graphs}
The line of thought which led us to the formulation of dimensional
concepts on networks and graphs in \cite{Req6} was essentially
motivated by physics. In statistical mechanics, for example, many
model systems are placed on a lattice being embedded in an ambient
euclidean space, $\R^d$. One observes that this embedding dimension,
$d$, then shows up as a relevant external parameter in many important
physical expressions. It governs phase transitions, critical behavior
and the decay properties of correlation functions, to mention only a
few examples. This restricted context however masks the intrinsic role
of something like dimension in these examples.

In a thought experiment we choose to ignore the embedding space and
treat the system as an array of interacting degrees of freedom sitting
on some labelled graph or discrete set of nodes. The physics has to
remain the same. We learn from this that what is really important is
the number of nodes or degrees of freedom a given fixed node can have
contact with after a certain sequence of consecutive steps on the
graph or sequence of elementary interactions.This process defines a
natural neighborhood structure, viz. fixes the intrinsic geometry of
the system.

We hence make the following series of definition.
\begin{defi}Let $B(x,r)$ be the ball of radius $r$ around vertex $x$
  on the graph $G$, viewed as a metric space with respect to the
  standard graph metric $d$ introduced in the introduction. That is
\begin{equation}y\in B(x,r)\;\text{if}\;d(x,y)\leq r
\end{equation}
Denote further by $\partial B(x,k)$ the set of vertices having exactly
the distance $k$ from vertex $x$ (which is for the above metric only
meaningful for $k\in \N_0$). By $|B(x,r)|\,,\,|\partial B(x,k)|$
respectively we designate the number of nodes lying in these sets.
\end{defi}
We now define \tit{growth function} and \tit{spherical growth
  function} on $G$ relative to some arbitrary but fixed vertex $x$.
(We use here the notation more common in geometric group theory. In
other fields it is also called the distance degree sequence, cf.
\cite{Harary}).
\begin{defi}The growth function $\beta(G,x,r)$ is defined by
\begin{equation}\beta(G,x,r)=|B(x,r)|    \end{equation}
Correspondingly we define
\begin{equation}\partial\beta(G,x,k):=\beta(G,x,k)-\beta(G,x,k-1)   \end{equation}
\end{defi}

With the help of the limiting behavior of $\beta\,,\,\partial\beta$ we
introduced two dimensional concepts in \cite{Req6}.
\begin{defi}The (upper,lower) internal scaling dimension with respect
  to the vertex $x$ is given by
\begin{equation}\overline{D}_s(x):=\limsup_{r\to\infty}(\ln\beta(x,r) 
/\ln r)\;,\,\underline{D}_s(x):=\liminf_{r\to\infty}(\ln\beta(x,r)/\ln r)
\end{equation}
The (upper,lower) connectivity dimension is defined correspondingly as
\begin{equation}\overline{D}_c(x):=\limsup_{k\to\infty}(\ln\partial\beta(x,k)
/\ln k)+1\;,\,\underline{D}_k(x):=\liminf_{k\to\infty}(\ln\beta(x,k)/\ln
k)+1
\end{equation}
If upper and lower limit coincide, we call it the internal scaling
dimension, the connectivity dimension, respectively.
\end{defi}
Remark:i) The two notions are not entirely the same in general whereas
they coincide for many models (this is quite similar to the many
different fractal dimensions). \\
ii) For regular lattices both yield the expected result, i.e. the
embedding dimension. In general however upper and lower limit are
different and non-integer. Similarities to fractal dimensions are not
accidental. For a more thorough discussion of all these points see
\cite{Req6}.\\
iii) These notions or similar ones have already been introduced earlier
(a point we were unaware of at the time of writing \cite{Req6}, see
for example \cite{Baxter}\cite{Dhar},\cite{Filk} or \cite{Scalettar}
and presumably elsewhere) but as far as we can see, such an idea was
either only mentioned in passing or its many interesting properties
never investigated in greater detail.\vspace{0.2cm}

It is important (in particular for physics) that these notions display
a marked rigidity against all sorts of deformations of the underlying
graph and are independent of the reference vertex for locally finite
graphs. We mention only two properties in this direction.
\begin{ob}i)If the vertex-degree of the graph is locally finite, the
  numerical values of the above quantities are independent of the
  reference vertex.\\
  ii)Insertions of arbitrarily many edges within a k-neigborhood of any
  vertex do not alter the dimension. Edge deletions, fulfilling a
  slightly more complicated locality property, do also not change
  these values (cf. lemma4.10 in \cite{Req6}, the discussion in
  sect.VII of \cite{Req2} and theorem 6.8 in \cite{Req3}). More
  specifically, edge deletions are called $k$-local if, in the
  transition from $G$ to $G'$, only edges are deleted in $G$ so that
  for the corresponding pairs of nodes, $(x,y)$, it holds that $y\in
  B_{G'}(x,k)$ with respect to $G'$.
\end{ob}

This geometric concept played an important role in our analysis of
ther discrete substructure of continuum space-time and a possible
limit behavior towards a continuous (smooth or fractal) macroscopic
geometry. It is therefore worth mentioning that similar concepts are
playing a crucial role in an important field of pure mathematics,
called \tit{geometric group theory} as we learned relatively recently
(cf. e.g. \cite{Harpe}). In this latter framework one typically studies
\tit{finitely generated} groups and visualize them as so-called
\tit{Cayley graphs}. For the record some brief definitions.
\begin{defi}The group $\Gamma$ is finitely generated, if there is a
  finite subset, $S\subset \Gamma$, such that every $g\in \Gamma$ can be
  represented as a finite word under group multiplication.
\begin{equation}g= s_1\cdots s_n \end{equation}
with $s_i\in S\cup S^{-1}=:S'\,,\,g^{-1}\in S^{-1}\;\text{if}\;g\in
S$. It is reasonable to exclude the unit element, $e$ from $S$.
\end{defi}
\begin{bem}It is important to note that $S\cap S^{-1}$ may happen to
  be non-empty! There may exist group elements with
  $g=g^{-1}$. For convenience one may assume $S^{-1}=S$ (i.e. $S$
  being inverse closed).
\end{bem}
We define a \tit{(word)metric} on $\Gamma$ in the following way. 
\begin{defi}Let $g,g'$ be represented by two words. Then the word
  metric $d_S$ is given by
\begin{equation}d_S(g,g'):=\inf_{|w|}|w|(g^{-1}\cdot g')=:l(g^{-1}\cdot
  g')\end{equation}
where $w(g^{-1}\cdot g')$ is a word representation of $g^{-1}\cdot g'$
with elements from $S'$, $|w|$ the length of the word and
the infimum is taken, as for graphs, over the length of the different
paths, connecting $g$ with $g'$. Evidently $l(g^{-1}\cdot g')$ is
simply the minimal word distance of $g^{-1}\cdot g'$ from
$e$ in the group, $\Gamma$.
\end{defi}
\begin{defi}The Cayley graph $C(S,\Gamma)$ has the elements
  $g\in\Gamma$ as vertices and an (unoriented) edge is drawn between
  $g,g'$ if $d_S(g,g')=1$, i.e., $g'=g\cdot s'\,,\,s'\in S'$.
\end{defi}
\begin{ob}With these definitions and with $e\not\in S$ the Cayley
  graph has no elementary loops or multi-edges and its vertex degree
  is regular ($v(g)=|S\cup S^{-1}|$).
\end{ob}
\begin{bem}The notion of Cayley graph may differ from author
  to author (cf. e.g. \cite{Harpe}, \cite{Godsil} or \cite{Bollo1}).
  With $S=S^{-1}=S'$ one may, for example, draw two oppositely
  oriented edges between nearest neighbor pairs $g,g'$, i.e.
  $g'=g\cdot s$, $g=g'\cdot s^{-1}$. Such a symmetric directed graph
  can however be associated with an undirected graph with vertex
  degree being half as large. In \cite{Bollo1} a Cayley graph is a
  oriented multi-graph. If it happens for some $s_1$ that
  $s_1=s_1^{-1}$, an edge may point for example from $e$ to $s_1$ and
  $s_1$ to $e$ (i.e. both edges coloured with $s_1$) while for
  $s_1\neq s_1^{-1}$ (and $s_1^{-1}\not\in S)$ $g$ and $gs_1$ are only
  connected by one oriented $s_1$-edge. As these algebraic details are
  not so important in our context, we deal in the following only with
  the unoriented Cayley graph defined above.
\end{bem}

We see that finitely generated groups fit nicely into our framework of
graph geometry, leading to a particular class of regular graphs.
Cayley graphs are however even more special. They represent a subclass
of the so-called \tit{vertex transitive} graphs, being examples of
\tit{homogeneous spaces}.
\begin{defi}A graph is vertex transitive if its automorphism group
  acts transitively, i.e. for all $x,y\in V$ there exists a graph
  automorphism, mapping $x$ to $y$. A graph automorphism is a
  bijective map
\begin{equation}F:(V,E)\to (V,E)     \end{equation}
which commutes with the incidence relation, i.e. $x$ is linked with
$y$ exactly if $Fx$ is linked to $Fy$.
\end{defi}
\begin{ob}Cayley graphs are vertex transitive.
\end{ob}
Proof: This follows directly from their definition. Group
multiplication from the left yields a subgroup of the automorphism
group and acts transitively.$\hfill \Box$\\[0.2cm]
Remark: Note that multiplication from the right is in general not an
isomorphism; it is a more general kind of transformation.\\[0.2cm]
This property is desirable from a physical point of view. It means
that the Cayley graph looks alike irrespectively of the reference
vertex being selected.

One should note that in general Cayley graphs are not uniquely related
to their groups. There usually exist different sets of generators
which generate the group. The respective Cayley graphs are hence in
general not isomorphic.  There exists, however, a much more natural
relation between them.
\begin{defi}Let $F$ be a map from a metric space, $X$, to a metric
  space, $Y$ with metrics $d_X,d_Y$. It is called a quasi-isometric
  embedding if the following holds: There exist constants,
  $\lambda\geq 1,\epsilon\geq 0$, such that
\begin{equation}\lambda^{-1}\cdot d_X(x,y)-\epsilon\leq
  d_Y(F(x),F(y))\leq \lambda\cdot d_X(x,y)+\epsilon \end{equation}
If, furthermore, there exists a constant $\epsilon'$ such that for all $y\in
Y$ we have $d_Y(y,F(X))\leq \epsilon'$, that is, $Y\subset
U_{\epsilon'}(F(X))$
 it is called a quasi-isometry; the
spaces are then called quasi-isometric. There is an equivalent
definition which shows that the preceding definition is in fact
symmetric between $X$ and $Y$ (see for example \cite{Harpe}). That is,
there exists a quasi-isometric map $G$ from $Y$ to $X$ with
corresponding constants and $d_X(G\circ F(x),x)\leq\rho$ and
$d_Y(F\circ G(y),y)\leq\rho$ for some $\rho$. If $\lambda=1$ it is
called a rough isometry.
\end{defi}
Remark: The latter statement can be proved by defining the inverse
map, $g:y\to g(y)\in X$, by selecting one of the possibly several
vertices, $x$, such that $g(y):=x$ with $d_Y(f(x),y)\leq \epsilon'$.\\[0.2cm] 
We then have
\begin{satz}On a finitely generated group, $G$, the word metric is
  unique up to quasi-isometry. That is, two finite sets of generators
  generate two Cayley graphs, which are quasi-isometric.
\end{satz}
Proof: The proof employs the fact that the generators of the one set
can be represented by words of finite length with respect to the other
set of generators.\bewende
\\[0.2cm]
Quasi-isometry is a very important concept and replaces the dull
category of isometric metric spaces (see the following sections). It
is in fact a much more natural concept in many respects in this wider
context.

In our investigation of the behavior of the dimension functions
$\overline{D}_s,\underline{D}_s$ or the growth series $\beta(x,k)$ in
our earlier work we found it surprisingly difficult to give general
characterizations of sufficiently large classes of graphs with, for
example, 
\begin{equation}\overline{D}<\infty\;,\;\overline{D}_s= \underline{D}_s=D_s\;,\;
D_s\in \N\quad\text{or}\quad\beta(x,k)\sim k^s\;\text{etc.}     \end{equation}
which, on the other hand, is very important if one wants to get a
sufficiently rich overview of continuum limit spaces figuring as
possible candidates of space-time continua. Possible relations to
fractal geometry were however discussed in section 5.2 of \cite{Req6}.

That it is difficult to give sufficient characterizations of large or
infinite graphs, having certain properties was already observed by
Erdos and Renyi at the end of the fifties of the last century and led
them to invent the field of \tit{random graph} theory (a standard
reference being \cite{Bollo2}). Furthermore, these problems are
related to the so-called \tit{word-problem} in combinatorial group
theory, which makes some of the roots of the problems perhaps better
understandable. We think, the close connections to the large and
florishing field of geometric group theory will be of quite some help
in this respect. In the rest of this section we report on some, in our
view, useful results being related to such issues. It will however
become apparent that our problems, reported above, are no accident.
Some of the theorems we will mention and discuss in the sequel are in
fact very deep, relatively recent and typically restricted to the
highly regular subclass of Cayley graphs or certain generalisations of
them.

The two extremes as to dimensional behavior are given by \tit{trees}
and \tit{regular lattices} respectively. For a regular infinite
tree of uniform vertex degree $v\geq 3$ we have for the growth series
relative to an arbitrary vertex $x$:
\begin{multline}\beta(0)=1\,,\,\beta(1)=1+v\,,\,\beta(2)=1+v(v-1)\,,\,\beta(k)=1+\sum_{\nu=0}^{k-1}(v-1)^{\nu}=\\
1+v\cdot\sum_{\mu=1}^k(v-1)^{\mu-1}=1+v\cdot[1-(v-1)^k]/[1-(v-1)]\sim
v^k
\end{multline}
for large $k$. This is clearly an exponential growth (for more details
concerning this notion see \cite{Harpe}). Our scaling dimension is
$\infty$ but one can define
\begin{equation}\omega(\Gamma):=\limsup_k(\beta(x,k,\Gamma))^{k^{-1}}
\end{equation}
as \tit{exponential growth type}.    

The other extreme is given by lattices like $\Z^n$ with 
\begin{equation}\beta(x,k,\Z^n)\sim A\cdot k^n     \end{equation}
and $D(\Z^n)=n$. Such graphs or groups are called being of
\tit{polynomial growth}.\\[0.2cm]
Remark: A little remark is in order here. Our graph metric corresponds
to the so-called $l_1$-metric in euclidean space, that is ($x\in\Z^n$)
\begin{equation}d_{l_1}(x,0)=\sum_{i=1}^n |x_i|    \end{equation}
Therefore the growth is not exactly the one found for the ordinary
euclidean metric. We have for example for $\Z^2$:
\begin{equation}\beta(k)=2k^2+2k+1\leq Ak^2    \end{equation}
for some $A$.\\[0.2cm]
\begin{defi}If $\beta(x,k,G)\lesssim k^{\overline{D}}$ for some
  $\overline{D}\geq 0$, the graph or group is called of polynomial
  growth. The degree of polynomial growth is then defined by 
 \begin{equation}\overline{D}(G)=\limsup_k \log\beta(k)/\log(k)       \end{equation}
Strictly speaking, polynomial growth in geometric group theory is
usually defined with an integer, $d$, in the exponent so that
$\overline{D}< d$ if it is not an integer.
\end{defi}
(Note that $\overline{D}$ is the same as our upper internal scaling
dimension we defined before we learned of the existence of the
parallel developements in geometric group theory).

It is important to note that quasi-isometric graphs, both having
globally bounded vertex degree, have the same growth type.
\begin{satz}\label{thquasi}Let $G_1,G_2$ be graphs with vertex degree globally
  bounded by $v$ and reference vertices $x_1,x_2$. Let $F$ be a
  quasi-isometric embedding of $G_1$ into $G_2$, then
\begin{equation}\label{quasi}\beta_1(k,x_1)\leq A\cdot\beta_2(\lambda k+b,x_2)       \end{equation}
with $b:=d_2(x_2,F(x_1))+\epsilon$. For $G_2$ having a polynomial growth degree
it follows that 
\begin{equation}\overline{D}_1(x_1)\leq
  \overline{D}_2(x_2)\;,\;\underline{D}_1(x_1)\leq
  \underline{D}_2(x_2)   \end{equation}
If $G_1,G_2$ are quasi-isometric these inequalities become equalities.
\end{satz}
Proof: Conceptually slightly different proofs of parts of the
statement can be found in
e.g. \cite{Req6}, \cite{Req2}sect.VII, \cite{Lochmann}, or
\cite{Harpe}. The notions and concepts, used in
\cite{Req6},\cite{Req2}sect.VII, are slightly different as at the time
of writing the papers we were not aware of the existing parallels in
geometric group theory. In \cite{Req2} we proved however a couple of
stronger results relating for example the growth function of a graph
to the growth function of the corresponding (coarse-grained)
\tit{clique-graph}. 

With the above defined ball, $B$, we have
\begin{equation}F(B(x_1,k))\subset B(x_2,\lambda k+B)       \end{equation}
Thus
\begin{equation}|F(B(x_1,k))|\leq\beta_2(x_2,\lambda k+b)       \end{equation}
If $F(x)=F(y)$, it follows from the property of quasi-isometric
embedding that $d_1(x,y)\leq \lambda\cdot\epsilon$. Due to the
globally bounded vertex degree there can be at most $A$ vertices in a
ball around $x$ with radius $\lambda\cdot\epsilon$. This can be easily
inferred from our estimates on trees by using for example a spanning
tree for the subgraph $B(x,\lambda\cdot\epsilon)$. Therefore
$B(x_1,k)$ can at most contain $A\cdot |F(B(x_1,k))|$ vertices,
yielding the estimate on the two growth functions. If $G_1,G_2$ are
quasi-isometric, such an estimate holds in both directions.

We come now to the estimates on the dimensions. For the $\limsup$ we
have the following. By assumption there exists a subsequence,
$k_{\nu}\to\infty$ so that 
\begin{equation}\ln\beta_1(x_1,k_{\nu})/\ln k_{\nu}\to
  \overline{D}_1(x_1)
        \end{equation}
Due to eqn(\ref{quasi}) and the properties of the logarithm the
corresponding subsequence with now $A\cdot\beta_2(\lambda k_{\nu}+b,x_2)$
inserted instead of $\beta_1(x_1,k_{\nu})$ is, on the one hand, an
upper bound on the previous subsequence and, on the other hand, has to
stay below $\overline{D}_2(x_2)$ assymptotically because of the
definition of $\limsup$. This proves the first statement. As to
$\liminf$, take now a subsequence, $r_{\nu}'$ so that
\begin{equation}\ln\beta_2(x_2,r_{\nu}')/\ln r_{\nu}'\to
  \underline{D}_2(x_2)            \end{equation}
We infer from eqn(\ref{quasi}) that
\begin{equation}\beta_1(r_{\nu},x_1)\leq A\cdot\beta_2(r_{\nu}',x_2)\end{equation}
with $r_{\nu}:=\lambda^{-1}(r_{\nu}'-B)\to\infty$ for
$r_{\nu}'\to\infty$. As before we conclude that
\begin{equation}\ln\beta_1(r_{\nu},x_1)/\ln r_{\nu}\leq
  \underline{D}_2(x_2)            \end{equation}
assymptotically. This then holds a fortiori for
$\underline{D}_1(x_1)$, yielding\\ the second statement.\bewende 

From the above we see that we have found a preliminary answer to one
point of our programme, i.e. finding classes of graphs or networks
having equivalent growth functions and the same dimensions.
\begin{ob}Quasi-isometry defines equivalence classes of graphs
  having the same dimension and equivalent growth functions in the
  category of graphs with globally bounded vertex degree.
\end{ob}
To get a better feeling to what extent quasi-isometry restricts the
structure of graphs, we show that a graph of globally bounded vertex
degree can even be roughly isometric to a graph with unbounded vertex
degree. 
\begin{ob}A graph with globally bounded node degree can be roughly
  isometric to a graph with unbounded node degree.
\end{ob}
Proof: Take for example some regular graph like $\Z^n$. Attach to each
node, $x$, or, more general, to a local neighborhood of such nodes, a
subgraph, $G_x$, such that all its nodes are linked to $x$, This
implies that the distance of two arbitrary nodes in $G_x$ as a
subgraph of this new graph, $G'$, have a distance $\leq 2$. As the
order of $G_x$ can be chosen arbitrary and dependent on $x$, one
easily gets such graphs as mentioned above by this construction. With
this $G'$ and the initial graph $G$ we can define the following map
\begin{equation}F:G'\ni x'\to x\in G    \end{equation}
where $x$ is the base-vertex in case $x'\in G_x$ or else the identity
map. We easily conclude that
\begin{equation}d(x',y')\leq d(x,y)+2\;,\;d(x',y')\geq d(x,y)   \end{equation}
so $F$ is evidently a rough isometry.\bewende

What remains is to find sufficiently large classes of graph
deformations which lead to quasi-isometries. In section 2 we
introduced the notion of $k$-local edge insertions and deletions. We
have already shown elsewhere that they do not alter the dimensions of
graphs without using explicitly the notion of quasi-isometry. (see the
references given there). To complete the discussion we now show that
they in fact lead to quasi-isometries.
\begin{propo}Let $G'$ arise from $G$ by a finite sequence of
  $k_{\nu}$-local edge insertions or deletions. All these
  transformations are quasi-isometries as is (because of transitivity)
  also the resulting operation. A fortiori, they even represent
  bilipschitz equivalences.
\end{propo}
\begin{defi}A map $F$ from the metric space $X$ to the metric space $X'$ is
$C$-lipschitz if $d'(Fx,Fy)\leq C\cdot d(x,y)$. It is a bilipschitz
equivalence if it is bijective and such inequalities hold in both
directions. This is equivalent to  
\begin{equation}C^{-1}\cdot d(x,y)\leq d'(Fx,Fy)\leq C\cdot d(x,y)
\end{equation}
for some positive $C$.
\end{defi}
\begin{ob} The usefulness of bilipschitz equivalences is that they
  also are topological homeomorphisms.
\end{ob}
Proof of the proposition: Let the vertices $x,x'$ denote the same
vertices in $G,G'$ respectively. Then $F:x\to x'$ is a bijective
map from $G$ onto $G'$ (more specifically, restricted to the
respective vertex sets). In the case of $k$-local edge insertions we
have
\begin{equation}k^{-1}\cdot d(x,y)\leq d'(Fx,Fy)\leq d(x,y)
\end{equation}
where only the lhs of the estimate is not entirely obvious. Take a
minimal path in $G'$ connecting $x'$ and $y'$ having length
$d'(x',y')$. It consists of a vertex sequence
$(x'=x'_0,x'_1,\ldots,y'=x'_d)$ with consecutive pairs having distance
equal to one. In $G$ this sequence corresponds to a sequence
$(x=x_0=x',x_1,\ldots,x_d=y=y')$. Due to the $k$-locality of edge
insertions, two consecutive vertices in the latter sequence can at
most be $k$ steps apart, i.e.
\begin{equation}d(x_i,x_{i+1})\leq k    \end{equation}
From this immediately follows that 
\begin{equation}d(x,y)\leq k\cdot d'(x',y')   \end{equation}
The case of $k$-local edge deletions is the inverse process if we
start from $G'$ and pass over to $G$ by $k$-local edge insertions. We
get
\begin{equation} d(x,y)\leq d'(x',y')\leq k\cdot d(x,y)   \end{equation}
That is, by symmetrizing we have the final estimate holding for both
insertions and deletions
\begin{equation}k^{-1}\cdot d(x,y)\leq d'(Fx,Fy)\leq k\cdot d(x,y)
\end{equation}
This completes the proof.\bewende
\section{The Case of Integer Dimension}
In \cite{Req6} we constructed examples of graphs having an arbitrary
real number as dimension. If dimension is a stable characteristic of
the limit process described in the following sections, such graphs
with non-integer dimension are expected to converge to some fractal
limit space. Motivated by physical applications, in particular in
foundational space-time physics, the case of integer dimension is of
course of particular importance. We know that for example $\Z^n$ has
integer graph-dimension $n$. Thus the preceding results guarantee that
local deformations of such lattice graphs yield again graphs with
integer dimension.
\begin{conclusion}Arbitrary k-local deformations of lattice graphs
  yield again graphs with integer dimension.
\end{conclusion}

We note that lattices like $\Z^n$ are Cayley graphs of finitely
generated abelian groups. The natural question is, are there other
classes of Cayley graphs having an integer dimension? Or more
generally, are there more general classes of (Cayley) graphs, having
polynomial growth? Abelian groups are a particular subclass of
so-called \tit{nilpotent} groups. We will not give the definition of
nilpotency here, which would need the introduction of some technical
machinery (see for example \cite{Bourbaki1}). Anyway, we have the
following deep theorem, which was proved in stages by several authors
(see \cite{Harpe} p.201).
\begin{satz}(Dixmier, Wolf, Guivarc'h, Bass) Let $\Gamma$ be a
  finitely generated nilpotent group. Then  $\Gamma$ is of polynomial
  growth . More precisely, there exist constants, $A,B$ and an integer
  $d$ so that
\begin{equation}Ak^d\leq\beta(k)\leq Bk^d   \end{equation}
\end{satz}
A stronger result was proved by Bass (\cite{Bass}).
\begin{satz}If the nilpotent finitely generated subgroup $H$
  has finite index in $\Gamma$, then $H$ and $\Gamma$ have the same
  integer growth degree. In that case one calls the group $\Gamma$
  almost nilpotent.
\end{satz}
This sequence of results culminated in the observation of Gromov:
\begin{satz}(Gromov)A finitely generated group has polynomial growth
  iff it contains a nilpotent subgroup of finite index (implying that
  the growth degree is an integer), see \cite{Gromov1}.
 \end{satz}
So we conclude that all $k$-locally deformed Cayley graphs of almost
nilpotent groups have integer dimension, which is an already quite
large class.

There exists an important and quite interesting example which is (for
various reasons) intensely studied in geometric group theory. It is
the so-called \tit{Heisenberg group}, $H$, (cf. e.g. \cite{Harpe}). Its
elements are triangular matrices of the form:
\begin{equation}g=\begin{pmatrix}1 & 0 & 0 \\
k & 1 & 0 \\ m & l & 1
\end{pmatrix}\;,\;k,l,m\in \Z        \end{equation}
with generators
\begin{equation}s= \begin{pmatrix}1 & 0 & 0 \\
1 & 1 & 0 \\ 0 & 0 & 1\end{pmatrix}\;,\;t=\begin{pmatrix}1 & 0 & 0 \\
0 & 1 & 0 \\ 0 & 1 & 1 \end{pmatrix}\;,\;u=\begin{pmatrix}1 & 0 & 0 \\0
& 1 & 0 \\1 & 0 & 1 \end{pmatrix} 
\end{equation}
and group laws
\begin{equation}su=us\;,\;tu=ut\;,\;ts=stu    \end{equation}
The general element with entries $k,l,m$ is the word $s^kt^lu^m$. So,
as a set, $H$ is standing in natural bijection to $\Z^3$, the latter
having dimension $3$. After some calculations one finds however
(\cite{Harpe}, p.197)
\begin{satz}For the Heisenberg group we have 
\begin{equation}Ak^4\leq\beta(k)\leq Bk^4  \end{equation}
i.e., its growth degree (or our internal scaling dimension) is four. 
\end{satz}
We conjecture that the reason for this surprising behavior derives
from one of the three generating relations, when representing $H$ as a
Cayley graph. The critical relation is
\begin{equation}(s^kt^lu^m)\cdot s= s^{(k+1)}t^lu^{(m+l)}     \end{equation}
(the other two behave smoothly as to the exponents, cf. \cite{Harpe}
p.198). That is, applying $s$ to a group element, $g$, may send it to
an element being far away from $g$ (for large $l$) in a naive metrical
$(\Z^3)$-sense.

We note that we studied possibly related phenomena in
\cite{Req2},\cite{Req3}, where we analysed the possibility that
space-dimension changes under, what we called, the coarse-graining or
renormalisation process. It turned out that only a very peculiar and
\tit{non-local} network wiring (with respect to some coarse-grained
metric), called by us \tit{critical network states}, allows for such a
dimensional reduction under renormalisation. This seems to be again a
point where both fields of research are closely related.

We close this section with another surprising and deep result of
advanced graph theory, having again certain ramifications into the
part of foundational space-time physics we described above. We already
remarked that important situations arise when the network under
investigation has an asymptotic dimension, $D$, and, a fortiori, if
this number $D$ is integer. We expect that by studying such particular
networks one can learn something useful about the reason why our
space-time also has an integer dimension at least on a macroscopic scale.

We have seen that Cayley graphs of polynomial growth have an integer
growth degree (which is quite a deep result), and that this also holds
for $k$-local deformations of such graphs and for graphs being
quasi-isometric to them. A natural class which fulfills a certain kind
of \tit{homogeneity} and comprises the Cayley graphs are the
so-called \tit{vertex transitive} graphs.
\begin{defi}A graph, $G$, is called vertex transitive if for each pair
  of vertices, $x,y$, there exists an element, $\alpha_{xy}$, in
  $AUT(G)$ with $\alpha_{xy}(x)=y$.
\end{defi}
The following is quite obvious.
\begin{propo}A vertex transitive graph has equal node degree for all
  nodes, $x$. Furthermore, the growth function, $\partial\beta(x,k)$,
  is independent of $x$.
\end{propo}
Proof: The first point follows directly from the definition. As to the
second property; for each pair of nodes, $x,y$, there exists by
assumption an automorphism, $\alpha$, with $\alpha(x)=y$ and
$\alpha^{-1}(y)=x$. Let $z$ be an element of $\partial B(x,k)$. There
exists a minimal path connecting $x$ and $z$. This path is mapped onto
a path of equal length, connecting $y$ and $\alpha(z)$. This path is
again minimal. To show that we assume that there exist another,
shorter, path connecting $y$ and $\alpha(z)$. It is mapped by
$\alpha^{-1}$ onto a path connecting $x$ and $z$ which is shorter than
the original minimal path, which is a contradiction. We hence have
\begin{equation}\alpha(\partial B(x,k))\subset \partial B(y,k)\;,\,
\;\alpha^{-1}(\partial B(y,k))\subset \partial B(x,k)   \end{equation}
and therefore
\begin{equation}\partial B(y,k)=\alpha\circ \alpha^{-1}(\partial
  B(y,k))\subset \alpha(\partial B(x,k))\subset \partial B(y,k)
\end{equation}
implying 
\begin{equation}\alpha(\partial B(x,k))= \partial B(y,k)       \end{equation}
This proves the proposition.\bewende  \vspace{0.2cm}

We saw that Cayley graphs are a true subclass of the class of vertex
transitive graphs, which, on their hand, are the type of networks we
would like to call \tit{strongly homogeneous}. As to this latter class
we have the following interesting theorem attributed to Sabidussi
(\cite{Sabidussi}) and discussed for example in \cite{Godsil}
sect. 3.9. 
\begin{satz}Any connected vertex transitive graph is a retract of a
  Cayley graph, where a subgraph, $Y$, of a graph, $X$, is called a
  retract of $X$ if there exists a graph-homomorphism, $f$, from $X$
  to $Y$ such that the restriction, $f|Y$, of $f$ to $Y$ is the
  identity map.
\end{satz} 
Remark: The construction of the Cayley graph from a vertex transitive
graph, $X$, essentially consists in replacing each vertex, $x$, by an
\tit{independent set} of vertices of size $|G_x|$, with $G_x$ the
stabilizer subgroup of $x$ in $AUT(X)$ and connecting these sets in a
bipartite way if the original vertices were neighbors in
$X$.\vspace{0.2cm}

If one can show that these local sets of vertices have finite
cardinality, one can, as in the situation of quasi-isometries or local
deformations, conclude that the vertex transitive graph has the same
dimension as the Cayley graph, so constructed. This was accomplished
in an ingeneous analysis by Trofimov (\cite{Trofimov}), see also the
nice survey by Imrich et al (\cite{Imrich}). An important role in this
context is played by the so-called \tit{systems of imprimitivity}. A
nice discussion of this concept can e.g. be found in \cite{Godsil}
sect.2.5.

The theorem reads:
\begin{satz}If the locally finite connected vertex transitive graph,
  $X$, has polynomial growth, the corresponding Cayley graph,
  described above is the graph of an almost nilpotent group of finite
  index. Hence its growth degree is an integer and the same holds for
  the vertex transitive graph $X$.
\end{satz}
We again conclude that this also holds for all the local deformations
of such graphs and graphs being quasi-isometric to them.
\section{The Gromov-Hausdorff Distance}
In \cite{Req1} and \cite{Req2} we developed a technical framework in
greater detail which we sketched already in earlier work. That is, we
invented a canonical process which allows us (at least in principle)
to construct a \tit{continuum limit space} from some underlying highly
irregular and erratic primordial substratum, which, we surmise, is
underlying our macroscopic space or space-time on the Planck scale
level. This represents however a formidable task in any approach to
quantum gravity which starts from some microscopic non-smooth
space-time. Therefore, several technical points of the procedure could
only be incompletely treated.

To tackle this problem on a sufficiently broad scale, i.e. avoiding
already from the beginning a too narrow starting point, we chose as
our model system, as in the previous sections, large irregular graphs
or networks, making the idealisation that the node sets are countably
infinite. 

Our central idea was it to distill in a sequence of systematic
\tit{coarse-graining} or \tit{renormalisation} steps some
\tit{large-scale} structure of our networks if there does exist
any. We in fact located specific criteria in our microscopic
substratum (some \tit{geometric criticallity} or \tit{long-range
  order}) which are expected to be crucial for the existence of an
interesting large-scale limit manifold. We do not intend to repeat a
description of this process in greater detail in the following, we
only want to make clear what our central goals are.

The process developed by us is reminiscent of the real-space
renormalisation process of the statistical mechanics of critical
systems. Instead of block-spins and regular lattices we select
certain densely entangled subgraphs (called cliques or lumps) in our
network and promote them to new nodes in a new (meta-)graph. We draw a
(meta-)edge between to cliques if they overlap (to a certain degree),
i.e. having a non-void set of common nodes. In this way we construct
the so-called \tit{clique-graph}.

One can now choose to forget about the internal structure of these new
nodes or average over the inner structure and repeat this process,
getting nodes and edges of the next higher level and so on. As in the
ordinary renormalisation process the idea is to compare these
structures living on the different scales of our graph with each
other, study their flow, hoping that they converge to a certain
\tit{fixed-structure}. It is evident that this process implements a
certain kind of coarse-graining, distilling possibly hidden
laqrge-scale characteristics of the graph or network. The main
technical tool in our analysis has been the theory of random graphs
(see for example \cite{Bollo2}). 

Performing this task we have to deal with formidable technical
problems. Note for example that it is far from clear under what
conditions we arrive at a smooth macroscopic manifold or, on the other
hand, a chaotic or fractal-like limit point. Furthermore, in order to
speak meaningful about concepts like limit or convergence,
topological, or even better, metrical concepts have to be introduced
or developed. These aspects will be discussed in greater detail in the
following.

The idea is to view space or space-time on several scales of
resolution at a time, from the very microscopic to the macroscopic
regime. In other words, we have to introduce a form of \tit{scaling
  limit} on graphs or metric spaces in general. It turns out that the
performance of such a general rescaling process leads to a whole bunch
of deep mathematical questions which have been treated only relatively
recently in pure mathematics (see for example \cite{Gromov},
\cite{Bridson}, \cite{Grigorchuk} or \cite{Semmes}).

What we need in the first place are fruitful notions of distance and
convergence in the category of general metric spaces, then leading to
further notions like e.g. completeness, compactness etc. in some
super-space of spaces. Such concepts have been developed by M.Gromov
and other people. What was in fact well-known is the notion of
\tit{Hausdorff-distance} of, for example, compact sets lying in some
ambient (compact) metric space.
\begin{defi}Let $X$ be a metric space, $U_{\epsilon}(A)$ the
  $\epsilon$-neigborhood of a subset $A\subset X$. The
  Hausdorff-distance between $A,B\subset X$ is then given by 
\begin{equation}d_H(A,B):=\inf\{\epsilon;A\subset
  U_{\epsilon}(B),B\subset U_{\epsilon}(A)\}  \end{equation}
\end{defi}
We have the following lemma
\begin{lemma}With $X$ a compact metric space, the closed subsets of
  $X$ form a compact (i.e. complete) metric space with respect to
  $d_H$ (see e.g. \cite{Bridson} or \cite{Edgar}).
\end{lemma}
In the following it is sometimes useful to make a slight
generalisation to pseudo metric spaces as we will encounter situations
where spaces or sets have zero Gromov-Hausdorff-distance (for example,
the one being a dense subset of the other) while they are not strictly
the same. Everything we will prove for metric spaces in the following
will also hold for pseudo metric spaces.
\begin{defi}A pseudo metric fulfills ther same axioms as a metric with
  the exception that $d(a,b)=0\,\to\,a=b$ does not necessarily hold.
\end{defi}

The above distance concept is too narrow to be useful in a more
general context. It was considerably generalized by M.Gromov in an
important way (see \cite{Gromov1}) and later slightly modified by
himself and other authors
(\cite{Gromov},\cite{Bridson},\cite{Petersen}). What is really
beautiful in our view is, that, while it seems to be more abstract, it
encodes the really important and crucial aspects of similarity or
``nearness'' of spaces in a more satisfying way. That is, it measures
their structural similarity and not simply the nearness of two
structureless sets of points in a space. In general it is a pseudo
metric which may even take the value infinity. For compact spaces it
is always finite. If one forms equivalence classes of compact spaces under
isometries, it becomes a true metric.

The Gromov-Hausdorff distance, $d_{GH}$, can be formulated in two
equivalent ways.
\begin{defi}$d_{GH}(X,Y)$ between two metric spaces, $X,Y$,is
  defined as the infimum of $d_H^Z(f(X),g(Y))$ over all metric spaces
  $Z$ and isometric embeddings, $f,g$, of $X,Y$ into $Z$.\\
Equivalently, one can define $d_{GH}$ by the infimum over $d_H(X,Y)$
in $X\sqcup Y$ equipped with the metric $d_{X\sqcup Y}$ which extends
the respective metrics $d_X,d_Y$ in $X,Y$.
\end{defi}
To give a certain impression how typical properties are proved in this
context we show that $d_{GH}$ fulfills the triangle inequality (the
proof of which is frequently scipped in the literature whereas it is
not entirely trivial, see also \cite{Lochmann}).

With $X,Y,Z$ metric spaces and (without loss of generality)
\begin{equation}d_{GH}(X,Y)<\infty\;,\;d_{GH}(Y,Z)<\infty
\end{equation}
we build the spaces
\begin{equation}X\sqcup Y\,,\,Y\sqcup Z\,,\,X\sqcup Z\,,\,X\sqcup
  Y\sqcup Z           \end{equation}
with $d_{X\sqcup Y},d_{Y\sqcup Z}$ metrics, extending
$d_X,d_Y,d_Z$. We define the following metric on $X\sqcup Z$, $X\sqcup
Y\sqcup Z$, respectively, extending the metrics $d_{X\sqcup
  Y},d_{Y\sqcup Z}$. 
\begin{equation}d^*(x,z):=\inf_y(d_{X\sqcup Y}(x,y),d_{Y\sqcup Z}(y,z))          \end{equation}

We first show that this defines a metric on $X\sqcup Y\sqcup Z$. The
critical property is, as always, the triangle inequality. Inserting
the definitions in, for example, the configuration
\begin{equation}d^*(x,z)+d^*(z,x')        \end{equation}
and regrouping the terms belonging to $X\sqcup Y,Y\sqcup Z$
respectively one gets 
\begin{multline}d^*(x,z)+d^*(z,x')=\inf_{y,y'}(d(x,y)+d(x',y')+d(y,z)+d(y',z))\\
\geq\inf_{y,y'}(d(x,y)+d(x',y')+d(y,y'))\geq d(x,x')  \end{multline}
Remark: Note that in general the supremum over
metrics is again a
metric. This does not hold in general for the infimum.\\[0.2cm]
One now has for the infimum over all admissible metrics:
\begin{equation}d_{GH}(X,Z)\leq \inf_{d^*} (d^*_H(X,Z))\leq \inf_{d^*}(d^*_H(X,Y)+d^*_H(Y,Z))        \end{equation}
But all the above particular metrics $d^*$ are fixed on $X\sqcup
Y,Y\sqcup Z$, being there $d_{X\sqcup Y},d_{Y\sqcup Z}$ respectively.
That is, $d^*_H(X,Y)=d^{X\sqcup Y}_H(X,Y)$ etc. Therefore we can take
on the rhs the infimum within the brackets over the respective
metrics on both spaces, $X\sqcup Y$ and $Y\sqcup Z$ independently!,
getting the final inequality for the GH-distance
\begin{equation}d_{GH}(X,Z)\leq  d_{GH}(X,Y)+ d_{GH}(Y,Z)     \end{equation}
thus proving the statement.\bewende

One has the following lemma.
\begin{lemma}Two compact spaces, $X,Y$, are isometric iff
  $d_{GH}(X,Y)=0$. That is, we have a true GH-metric when taking
  isometry classes (\cite{Bridson}, p.73).
\end{lemma}
We have the further result
\begin{propo}The space $\mcal{C}$ of compact metric spaces is complete
  under $d_{GH}$ (see \cite{Petersen}).
\end{propo}
Remark: It is usually difficult to calculate the GH-distance
exactly. In many cases it is however sufficient to find good upper
bounds. In convergence questions this is frequently relatively
easy.\vspace{0.2cm}

In the following the concept of an $\epsilon$-net will become a useful
tool.
\begin{defi}An $\epsilon$-net in a metric space $X$ is a subset, $S$,
  which is $\epsilon$-dense in $X$, i.e. with $x\in X$ it follows that
  $d(x,S)\leq\epsilon$. Or, stated differently, the union of
  $\epsilon$-balls with centers in $S$ covers $X$.
\end{defi}
Of particular importance are finite $\epsilon$-nets.
\begin{defi}$X$ is called totally bounded if for all $\epsilon$ there
  exists an $\epsilon$-net which is finite.
\end{defi}
As a consequence we have
\begin{lemma}A metric complete space is compact iff it is totally
  bounded.
\end{lemma}
\begin{koro}A compact space has a finite diameter.
\end{koro}
\begin{koro}If $X,Y$ are compact metric spaces, a metric on $X\sqcup
  Y$ is for example defined by an extension of $d_X,d_Y$ in the form
\begin{equation}d_{X\sqcup Y}(x,y):=1/2\max\{diam\,X,diam\,Y\}=:1/2D
\end{equation}
It follows
\begin{equation}d_{GH}(X,Y)\leq 1/2D      \end{equation}
\end{koro}
Remark: Usually the crucial and sometimes somewhat tricky point in
finding new metrics or extending given metrics is to fulfill the
triangle inequality for all! different possible configurations. In the
above example this was, somewhat untypically, relatively
simple.\vspace{0.2cm}

To exhibit more clearly that a small GH-distance says something about
a certain metric similarity between general metric spaces, we discuss
the relation between rough isometry and GH-distance. This is
implicitly discussed in example 2 p.491 in \cite{Petersen}, the
discussion following definition 5.33 in \cite{Bridson} (where
GH-distance is introduced via so-called $\epsilon-relations$) or
proposition 5.3 in \cite{Gromov}. We state the following result as
formulated in \cite{Lochmann}.
\begin{satz}Two metric spaces have finite GH-distance iff they are
  roughly isometric.
\end{satz}
Proof: We assume that $d_{GH}(X,Y)<\epsilon$. Then, to each $x\in X$
there exists an $y$ so that $d(x,y)<\epsilon$ in an appropriate metric
$d$ on $X\sqcup Y$. We define a map
\begin{equation} f:x\to f(x)\in Y      \end{equation}
by selecting one of these elements $y$.  The triangle inequality
yields for example for the following configuration
\begin{equation}d(x,x')\leq d(x,f(x))+d(f(x),f(x'))+d(f(x'),x')\leq
  d(f(x),f(x'))+2\epsilon     \end{equation}
A corresponding relation holds in the other direction with a map
$g:y\to g(y)\in X$. Furthermore, we have $f\circ g(y)\in Y$ with
\begin{equation} d(y,f\circ g(y))\leq d(y,g(y))+d(g(y),f\circ
  g(y))\leq 2\epsilon   \end{equation}
We hence conclude that 
\begin{equation}d_Y(y,f(X))\leq 2\epsilon   \end{equation}
 for all $y\in Y$. This proves that $f,g$ define rough isometries.
\begin{koro}Furthermore, for the in general non-unique map, $x\to
  f(x)$ with $d(x,f(x))<\epsilon$, we have with another choice, $x\to
  f'(x), d(x,f'(x))<\epsilon$, that $d(f(x),f'(x))<2\epsilon$. Hence
  all points in $Y$ which are close to $x$ are also close to each
  other.
\end{koro}
Proof of corollary: 
\begin{equation}d(f(x),f'(x))\leq d(f(x),x)+d(x,f'(x))\leq 2\epsilon     \end{equation}

To prove the other direction in our theorem we introduce, similar to
\cite{Petersen} eample 2 or \cite{Gromov} proposition 3.5, a metric
$d$ on $X\sqcup Y$, exploiting the rough isometry of $X,Y$ with
parameters $\epsilon,\epsilon'$, i.e.
\begin{equation}d(x,x')-\epsilon\leq d(f(x),f(x'))\leq d(x,x')+\epsilon     \end{equation}
implying 
\begin{equation}|d(f(x),f(x'))- d(x,x')|\leq\epsilon     \end{equation}
and $Y\subset U_{\epsilon'}(f(X))$ with a corresponding relation in
the other direction.

We have seen that, with $\rho:=\max(\epsilon,\epsilon')$, $f(X)$ is an
$\rho$-net in $Y$ and that, as a consequence of the preceding
equation, $\{(x,f(x))\}$ defines a $\rho$-relation (cf. \cite{Bridson}
definition 5.33). Then by
\begin{equation}d(x,y):=\inf_{x'\in X}(d_X(x,x')+d_Y(f(x'),y)+\rho)       \end{equation}
a metric is defined on $X\sqcup Y$ extending $d_X,d_Y$. To illustrate
this point we show that the triangle inequality holds for the
following configuration
\begin{multline}d(x_1,y)+d(y,x_2)=\inf_{x',x''}(d(x_1,x')+d(f(x'),y)+d(x_2,x'')+d(f(x''),y)+2\rho)\\
\geq\inf_{x',x''}(d(x_1,x')+d(x_2,x'')+d(f(x'),f(x''))+2\rho)
 \end{multline}

From $f$ being a rough isometry we infer
\begin{equation}d(f(x'),f(x''))\geq d(x',x'')-\rho     \end{equation}
and hence
\begin{multline}d(x_1,y)+d(y,x_2)\geq\inf_{x',x''}(d(x_1,x')+d(x',x'')+d(x'',x_2)+\rho)\\
\geq d(x_1,x_2)+\rho>d(x_1,x_2)
\end{multline}
For $y:=f(x)$ we have $d(x,f(x))\leq\rho$, so $X$ lies in the
$U_{\rho}$-neigborhood of $Y$ with respect to this metric. On the
other hand we have for all $y$ 
\begin{equation}d(x,y)=\inf_{x'}(d(x,x')+d(f(x'),y)+\rho)\leq d(f(x),y)+\rho    \end{equation}
( by inserting $x$ for $x'$). Due to the assumptions there exists an
$x$ so that $d(f(x),y)\leq\rho$. This proves the theorem.\bewende
\begin{ob}Note that the set $\{(x,f(x))\}\cup \{(g(y),y)\}$ define a
  surjective $2\rho$-relation in the sense of \cite{Bridson} definition
  5.33, by means of which the GH-distance is defined there.
\end{ob}
Proof: This follows from $d_Y(f\circ g(y),y)\leq\rho$ (cf. the
definition of quasi-isometry in section 2) and
\begin{equation}d_Y(f(x),y)\leq d_Y(f(x),f\circ g(y))+d_Y(f\circ
  g(y),y)\leq d_X(x,g(y))+\rho+\rho   \end{equation}
and hence
\begin{equation}|d_X(x,g(y))-d_Y(f(x),y)|\leq 2\rho   \end{equation}

We now present the fundamental Gromov-compactness theorem, first for
compact spaces, then for more general cases.
\begin{defi}We call a family of compact spaces, $X_{\lambda}$, uniformly
  compact if their diameters are uniformly bounded and if for each
  $\epsilon>0$ $X_{\lambda}$ is coverable by $N_\epsilon<\infty$ balls
  of radius $\epsilon$ independent of the index $\lambda$.
\end{defi}
\begin{satz}(Gromov) A sequence $\{X_i\}$ contains a convergent
  subsequence in $d_{GH}$ iff $\{X_i\}$ is uniformly compact.
\end{satz}
Proof: see \cite{Gromov1},\cite{Petersen} or \cite{Bridson}. Typically
an \tit{Arzela-Ascoli-Cantor-diagonal-sequence}-like argument is used
in the proof.\bewende

In our framework we are mainly interested in infinite graphs,
i.e. non-compact metric spaces being however frequently \tit{proper}.
\begin{defi}A metric space, $X$, is called proper if all its closed
  balls, $B(x,r)$, are compact.
\end{defi}
We can then extend the above result in the following way. Ordinary
GH-convergence works well in the category of compact metric spaces. If
the spaces are non-compact, a slightly modified approach is more
satisfactory. One problem which may arise is that things in unbounded
spaces can ``wander away'' to infinity. So it is reasonable to pin
down the members of the sequence of spaces at certain points, so that
they can be better compared. More precisely, we work in the category
of \tit{pointed} metric spaces, $(X,x)$, which is, a fortiori pretty
normal from the physical point of view as it is like introducing a
reference point or a coordinate system.
\begin{defi}The sequence of pointed metric spaces, $(X_i,x_i)$, is
  said to converge to $(X,x)$ in pointed GH-sense if for every $r>0$
  the sequence of closed balls, $B(x_i,r)$, converges to $B(x,r)$ in
  $d_{GH}$.
\end{defi}
The Gromov-uniform-compactness theorem now reads:
\begin{satz}If for all $r$ and $\epsilon>0$ the balls $B(x_i,r)$ of a
  given sequence $(X_i,x_i)$ are
  uniformly compact, then a subsequence of spaces converges in pointed GH-sense.
\end{satz} 
Remark: There exist various slightly different notions of pointed
convergence in the literature. One can for example define pointed
GH-distance by admitting only isometries which map the base points
onto each other (\cite{Gromov}). Another possibility is to include the
distance of the images of the base points in the definition (\cite
{Petersen2}). The above definition is used in \cite{Bridson}.

\section{The Scaling-Limit of Infinite Graphs}
We now apply the techniques and results, developed in the preceding
sections, to our original problem, namely, investigating a particular
kind of scaling-limit in our space of metric spaces and apply it to
the subclass of locally bounded infinite graphs.

We start with a graph, $G$, of globally bounded vertex degree, $v$,
and, taking $G$ with the original graph metric, $d$, as initial metric
space, generate a sequence, or, more generally, a directed system of
metric spaces, $\lambda G$, by taking the same graph, $G$, but now
with the scaled metric, $\lambda d$, defined as
\begin{equation}\lambda d(x,y):=\lambda\cdot d(x,y)  \end{equation}
and (usually) taking $\lambda\to 0$. One may, in particular, take
subsequences of the kind
\begin{equation}G_n\;,\;d_n:=n^{-1}\cdot d\;,\;n\to\infty       \end{equation}
or replace $n$ by $2^{-k}$.

In a first step we have to provide criteria under which the balls,
$B_n(x,r)\subset G_n$, with $x$ a fixed reference vertex in $G$, are
uniformly compact. To this end we have to introduce a new concept. 
\begin{defi}With $(X,d)$ a metric space, $\mu$ a positive Borel
  measure on $X$, $\mu$ is said to be doubling, if there exists a
  positive constant, $C$, being independent of $B$ such that
\begin{equation}\mu(2B)\leq C\cdot\mu(B)       \end{equation}
for all balls in $X$. Note that $2B$ is a ball with the same center as
$B$ but twice the radius.
\end{defi}
\begin{lemma}It easily follows via iteration that
\begin{equation}\mu(2^kB)\leq C^k\cdot\mu(B)   \end{equation}
\end{lemma}
\begin{defi}A metric space is called doubling if each ball, $B$, can
  be covered by at most $C$ balls with radius half that of $B$ with
  again $C$ independent of $B$.
\end{defi}
\begin{propo}If $(X,d)$ has a doubling measure, it is doubling as a
  metric space.
\end{propo}
Proof: See \cite{Gromov}, p.412 (the chapter being written by
Semmes).\vspace{0.2cm}

Now take in the category of metric spaces the notion of \tit{pointed
  convergence} and apply it to the sequence of spaces
$X_n:=(X,d_n)$. The ball $B_n(x,r)$ in $X_n$ corresponds as a set to the
ball $B(x,nr)$ in $X$. The sequence of balls $B_n(x,r)$ is uniformly
compact if each ball can be covered by at most $N_{\epsilon}(r)$
$\epsilon$-balls. This means, that all $B(x,nr)$-balls can be covered
by at most $N_{\epsilon}(r)$ balls of radius $n\cdot \epsilon$ in
$X$. This is the case if $X$ carries a doubling measure $\mu$
according to the preceding lemma.
So what we have to analyse is, under what conditions do graphs cary a
doubling counting-measure with
\begin{equation}\mu(B(x,k)):=|B(x,k)|   \end{equation}

To this end we have to formulate in the general case (i.e. graphs not
necessarily being Cayley or vertex-transitive graphs) a slightly more
restricted form of polynomial growth.
\begin{defi}We say, a locally finite graph, $G$, has uniform
  polynomial growth if there exist constants, $A,B,d>0$ so that 
\begin{equation}Ak^d\leq \beta(x,k)\leq Bk^d   \end{equation}
for all $k\geq k_0$ and $A,B$ independent of the reference point $x$.
\end{defi}
Remark: It is interesting that the dimension, $d$, is independent of the
reference point in a locally finite graph but to show the doubling
property, one needs a stronger result, i.e. the constant $A$ has to
stay away from zero if $x$ varies in $V(G)$. This is another weak form
of homogeneity of a graph.\vspace{0.2cm}

If our graph, $G$, has uniform polynomial growth one has the following
estimate for sufficiently large $k$:
\begin{equation}\beta(x,2k)\leq B\cdot 2^d\cdot k^d\;,\;\beta(x,k)\geq
  A\cdot k^d   \end{equation}
hence
\begin{equation}\beta(x,2k)\leq B/A\cdot 2^d\cdot\beta(x,k)       \end{equation}
\begin{conclusion}A graph with uniform polynomial growth has a
  doubling counting measure for sufficiently large $k$ and is hence
  doubling as a metric space for sufficiently large $k$. This implies
  that all balls $B_n(x,r)$ in $G_n$ are uniformly compact.
\end{conclusion}
The question is now, which graphs have this property of weak
homogeneity? It is clear that locally finite vertex-transitive graphs
of polynomial growth have this property. 

We assume the following. Let $G_1$ have uniform polynomial growth for
sufficiently large $r\geq r_0$, i.e.
\begin{equation}C_1\cdot r^d\leq\beta(x,r)\leq C_2\cdot r^d      \end{equation}
Assume, furthermore, that the graphs $G_1,G_2$ have globally bounded
vertex degree and that they are quasi-isometric. For convenience we
choose all ocurring constants to be symmetric with respect to the
quasi-isometric maps, $f:G_1\to G_2\;,\;g:G_2\to G_1$ (cf. the
definition of quasi-isometry in section 2, we assume also
$\epsilon=\epsilon'=\rho$). 

We know that $f(G_1)$ is $\rho$-dense in $G_2$ and vice versa. Taking
a ball, $B(y,r)$, around a vertex $y\in G_2$, there exists by
assumption a vertex $x\in G_1$ with $d_2(f(x),y)\leq\rho$.
We have shown (see the first part of theorem (\ref{thquasi})) that 
\begin{equation}\beta_1(x,r)\leq A\cdot \beta_2(y,\lambda r+2\rho)  \end{equation}
with $A$ independent of $x,y$ (for globally bounded vertex degree). We
hence have  
\begin{equation}\beta_2(y,\lambda r+2\rho)\geq A^{-1}C_1r^d     \end{equation}
Correspondingly we have
\begin{equation}A\beta_1(x,\lambda r+2\rho)\geq\beta_2(y,r)    \end{equation}
Now employing the universal polynomial growth for $G_1$ we get
\begin{equation}A^{-1}C_1(\lambda^{-1}(r-2\rho))^d\leq\beta_2(y,r)\leq
AC_2(\lambda r+2\rho)\end{equation}
Inserting for $r$ either $r$ (large enough) or $2r$ we arrive at
\begin{equation}\beta_2(y,2r)\leq AC_2(2\lambda
  r+2\rho)^d\;,\;\beta_2(y,r)\geq A^{-1}C_1(\lambda^{-1}(r-2\rho))^d    \end{equation}
and therefore
\begin{equation}\beta_2(y,2r)/\beta_2(y,r)\leq A^2\cdot
  C_2/C_1\cdot\lambda^d\cdot \frac{(2\lambda r+2\rho)^d}{(r-2\rho)^d}    \end{equation}
For $r$ sufficiently large (e.g. $2\rho\leq r/2$) we can bound the rhs
by a constant from above. We thus have
\begin{satz}Let $G_1,G_2$ both have globally bounded vertex degree,
  let $G_1$ have uniform polynomial growth, and let $G_1,G_2$ be
  quasi-isometric. Then also $G_2$ has uniform polynomial growth. We
  conclude that graphs with this property have a scaling limit in the
  sense discussed above.
\end{satz} 

We can make use of the theorem in the following way. We know that
locally finite vertex transitive graphs of polynomial growth have, by
the same token, a uniform polynomial growth. So all graphs of globally
bounded node degree, being quasi-isometric to such graphs, qualify as
candidates having a continuum limit under scaling. A fortiori we know
that they even have an integer dimension. That is, we can dispose of a
quite large class of physically interesting examples. 

Furthermore, our approach shows that there are certain similarities to
\tit{attractors} in dynamical systems. We have in fact a kind of
\tit{universality}.
\begin{koro}Let $G$ have the scaling limit $X$. Let $G'$ have finite
  GH-distance to $G$. It follows that $G'$ also converges toward $X$
  in $d_{GH}$.
\end{koro}
Proof: Let $d_{GH}(G,G')\leq a$. The relations between the metrics $d$
and the scaled metrics $\lambda\cdot d$ are bijective. It is hence easy
to show that it follows that $d_{GH}(G_n,G'_n)\leq n^{-1}\cdot a$ and
therefore
\begin{equation}d_{GH}(X,G'_n)\leq d_{GH}(X,G_n)+d_{GH}(G_n,G'_n)
\end{equation} 
which proves the statement.\bewende

A last point we want to discuss is the following. From the point of
view of physics an analysis of the properties of the continuum limit
space, $X$, as a consequence of certain characteristics of the
initial space, $G$, is extremely important. One can in fact prove
quite a lot in this respect. We restrict ourselves, for the sake of
brevity, in this paper to showing the following remarkable property.

Let the graph, $G$, be of uniform polynomial growth (growth degree
$d$). We learned that under this condition its natural counting
measure is doubling which implies that $G$ as a metric space is
doubling. We know in fact a little bit more. Under the mentioned
conditions all balls, $B(x,r)$, in $G$ are totally bounded viz.
compact and the number of $\epsilon$-balls, $N_{\epsilon}(r)$, needed
to cover them can be easily estimated with the help of our previous
results. It follows from our assumptions that $N_{\epsilon}(r)$ scales
with $(\epsilon,r)$ in the form
\begin{equation}N_{\epsilon}(r)\sim (r/\epsilon)^d      \end{equation} 
Note that in the following we need this property only for
large $n$, i.e. balls, $B_n(r)$, in $G_n$ which correspond to balls
$B(nr)$ in $G$ and, hence, $\epsilon$-balls in $G_n$ corresponding to
$n\epsilon$-balls in $G$.

That is, in order to be allowed to choose $\epsilon$ arbitrarily small
and still having non-trivial $\epsilon$-balls in the $G_n$, we can
always assume $n$ to be sufficiently large. For convenience we ignore
this technical point in the following. The same scaling as above holds
for the minimum, $cov_{\epsilon}(B_r)$, of such $N_{\epsilon}(r)$.
That is, we have
\begin{lemma}For sufficiently large $n$ we have
\begin{equation}cov_{\epsilon}(B_r)\sim (r/\epsilon)^d
\end{equation}
the $d$ coming from the uniform polynomial growth.
\end{lemma}

We now choose $n$ so large that
$d_{GH}(B_X(x,r),B_n(x,r))\leq\rho$. From our previous results we then
have rough isometries $f,g$ between $B_X(x,r),B_n(x,r)$ with
$d(f(B_n),y)\leq 2\rho$ for all $y\in B_X(x,r)$ and the same result in the
other direction. We select a minimum number,
$cov_{\epsilon}(B_n(x,r))$, of points $x_i\in B_n(x,r)$, so that $B_n(x,r)$ 
is covered by $\epsilon$-balls centered at $x_i$. Each $y\in B_X(x,r)$ has
distance at most $2\rho$ to an $\epsilon$-ball centered at some
$f(x_i)$. Hence, for all such $y$ there exists a $x_i$ so that
\begin{equation}d_X(y,f(x_i))\leq \epsilon+2\rho   \end{equation}
The same holds for the opposite direction and the map $g:X\to G_n$. As
the number of points $x_i$ was chosen minimal by assumption, we
conclude
\begin{propo}We have
\begin{equation}cov_{\epsilon+2\rho}(B_X(x,r))\leq
  cov_{\epsilon}(B_n(x,r))\;,\;cov_{\epsilon+2\rho}(B_n(x,r))\leq cov_{\epsilon}(B_X(x,r))   \end{equation}
and thus
\begin{equation}cov_{\epsilon+2\rho}(B_n(x,r))\leq
  cov_{\epsilon}(B_X(x,r))\leq cov_{\epsilon-2\rho}(B_n(x,r))  \end{equation}
or
\begin{equation}(r/\epsilon+2\rho)^d\leq cov_{\epsilon}(B_X(x,r))\leq (r/\epsilon-2\rho)^d  \end{equation}
\end{propo}

We can now choose $n$ large enough. This allows us to choose
$\epsilon$ arbitrarily small. For $\rho\to 0$ and then $\epsilon\to 0$
we get
\begin{satz}Let $G$ be of uniform polynomial growth with growth degree
  $d$. Then for the balls, $B_X(x,r)$, in the limit space X
 it holds
\begin{equation}cov_{\epsilon}(B_X(x,r))=\lim_{n\to\infty}cov_{\epsilon}(B_n(x,r))  \end{equation}
with the same scaling degree $d$. Furthermore, all balls in $X$ are
totally bounded, hence compact. Taking the infimum over $\epsilon$ we
call $d$ the covering dimension of $X$.
\end{satz}
As the covering dimension is related to the Hausdorff-measure, this
result shows, that also as measure spaces the $G_n$ converge toward a
reasonable continuum limit. We note that in this connection a lot more
can actually be shown but we choose to stop here.
\section{A Brief Outlook}
We have shown in the preceding sections that under certain assumptions
the existence of a macroscopic continuum limit can be guaranteed when
we start from an underlying presumably quite erratic substratum which
we modelled as a discrete metric space consisting of elementary
degrees of freedom and their elementary interactions or relations. It
turned out that the characteristics to be imposed on this primordial
network were reasonable from a physical point of view. A particularly
important notion was the concept of an intrinsic general dimension of
arbitrary discrete spaces. We could show that there even exist
criteria such that this dimension takes on integer values.

It is of great importance to learn under what conditions this limit
space is a smooth manifold or, on the other hand, a chaotic space of
rather fractal type. A particularly important possibility is a space
having a superficially smooth structure together with an internal
infinitesimal more erratic structure ``around'' the ``classical''
points of the base manifold, being kind of a generalized fiber space.
In this context a lot more can be investigated as e.g. (functional)
analysis, field theory or algebraic structures on the discrete spaces
and their respective continuum limits. We indicated how this works in
the case of measure theory.

As a last remark, there exists a slightly different approach developed
in \cite{Semmes} which we only briefly mentioned. It makes use of the
possibility of embedding the general spaces into some $\R^n$ (and
relies on several deep theorems). This approach, while perhaps being
more useful from a practical point of view, as it is frequently easier
to formulate concepts and prove results in $\R^n$, is, on the other
hand, slightly less general. So we preferred to begin our analysis
within the presumably more general framework. Nevertheless we hope to
come back to this latter lines of reasoning in the future.
\\[1cm]
{\small Acknowledgement: Discussions with A.Lochmann are gratefully
  acknowledged.}


\begin{thebibliography}{99}
  {\small
\bibitem{Berger}M.Berger:``Encounter with a Geometer, I,II'', Notices
of the AMS 47.2,3(2000)p.183,326
\bibitem{Markopoulou}F.Markopoulou: ``An Algebraic Approach to Coarse
  Graining'', hep-th/0006199
\bibitem{Oeckl}R.Oeckl: ``Renormalisation of Discrete Models without
  Background'', gr-qc/021204
\bibitem{Seriu}M.Seriu: ``Space of Spaces as a Metric Space'',
Comm.Math.Phys. 209(2000)393
\bibitem{Isham}C.Isham: ``Quantum norm theory and the quantisation of
  metric topology'', Class.Quant.Grav. 7(1990)1053
\bibitem{Req1}M.Requardt: ``(Quantum) Space-Time as a Statistical
  Geometry of Lumps in Random Networks'',
  Class.Quant.Grav. 17(2000)2029, gr-qc/9912059
\bibitem{Req2}M.Requardt: ``A Geometric Renormalisation Group and
  Fixed Point Behavior in Discrete Quantum Space-Time'', JMP 44(2003)5588,
  gr-qc/0110077
\bibitem{Req3}M.Requardt:``Wormhole Spaces, Connes' ``Points Speaking to
  Each Other'', and the Translocal Structure of Quantum Theory'',
  hep-th/0205168 (v.3 March 2004)
\bibitem{Borchers1}H.-J.Borchers,R.N.Sen: ``Theory of Ordered Spaces'',
  Comm.Math.Phys. 132(1990)593
\bibitem{Borchers2}H.-J.Borchers,R.N.Sen: ``Theory of Ordered Spaces,
  II The Local Differential Structure'', Comm.Math.Phys. 204(1999)475
\bibitem{Req4}M.Requardt:``Supersymmetry on Graphs and Networks'',
  Int.J.Geom.Meth.Mod.Phys. 2(2005)585, math-ph/0410059
\bibitem{Bollo1}B.Bollobas: ``Modern Graph Theory'', Springer, Berlin 1998
\bibitem{Req5}M.Requardt: ``Dirac Operators and the Calculation of
  the Connes Metric on arbitrary (Infinite) Graphs'',
  J.Phys.A:Math.Gen. 35(2002)759, math-ph/0108007
\bibitem{Req6}T.Nowotny,M.Requardt: ``Dimension Theory on Graphs and
  Networks'', J.Phys.A:Math.Gen. 31(1998)2447, hep-th/9707082
\bibitem{Harary}F.Buckley,F.Harary: ``Distance in Graphs'',
  Addison-Wesley, N.Y. 1990
\bibitem{Harpe}P.de la Harpe: ``Topics in Geometric Group Theory'',
  Univ. Chicago Pr., Chicago 2000
\bibitem{Bourbaki1}N.Bourbaki: ``Algebra'', chapt. I.6, Springer,
  Berlin 1991 
\bibitem{Bass}H.Bass: ``The Degree of Polynomial Growth of Finite
  Growth Nilpotent Groups'', Proc.London Math.Soc. 3Ser. 25(1972)603
\bibitem{Godsil}C.Godsil,G.Royle: ``Algebraic Graph Theory'',
  Springer, Berlin 2001
\bibitem{Bollo2}B.Bollobas: ``Random Graphs'' Sec.Ed., Cambridge
  Univ.Pr., Cambridge 2001
\bibitem{Baxter}R.J.Baxter:``Exactly solvable Models in Statistical
  Mechanics``, Academic Pr., N.Y. 1982
\bibitem{Dhar}D.Dhar: ``Lattices of Effectively Nonintegral
  Dimensionality'', J.Math.Phys. 18(1977)577
\bibitem{Filk}Th.Filk: ``Equivalence of Massive Propagator Distance
  and Mathematical Distance on Graphs'', Mod.Phys.Lett. A7 (1992)2637
\bibitem{Scalettar}R.T.Scalettar: ``Critical Properties of an Ising
  Model with Dilute Long Range Interactions'', Physica A170 (1991)282
\bibitem{Sabidussi}G.Sabidussi: ``Vertex-transitive graphs'',
  Monatshefte Math. 68(1964)427
\bibitem{Imrich}W.Imrich,N.Seifter: ``A survey on graphs with
  polynomial growth'', Discrete Math. 95(1991)101
\bibitem{Trofimov}V.I.Trofimov: ``Graphs with Polynomial Growth'',
  Math.USSR Sbornik 51(1985)405
\bibitem{Gromov}M.Gromov: ``Metric Structures for Riemannian and
  Non-Riemannian Spaces'', Birkhaeuser, N.Y. 1998
\bibitem{Bridson}M.R.Bridson,A.Haeflinger: ``Metric Spaces of
  Non-Positive Curvature'', Springer, N.Y. 1999
\bibitem{Grigorchuk}L.Bartholdi,R.Grigorchuk,V.Nekrashevych: ``From
  Fractal Groups to Fractal Sets'', math.GR/0202001
\bibitem{Semmes}G.David,S.Semmes:``Fractured Fractals and Broken
  Dreams, Selfsimilar Geometry through Metric and Measure'', Oxford
  Univ.Pr., Oxford 1997
\bibitem{Edgar}G.A.Edgar:``Measure,Topology, and Fractal Geometry'',
  Springer, Berlin 1990
\bibitem{Gromov1}M.Gromov:``Groups of Polynomial Growth and Expanding
  Maps'', Publ.Math.IHES 53(1981)53
\bibitem{Petersen}P.Petersen:``Gromov-Hausdorff Convergence of Metric
  Spaces'', AMS Proc.Pure Math. 54,3(1993)489
\bibitem{Petersen2}P.Petersen: ``Riemannian Geometry'', chapt.10,
  Springer, Berlin 1991
\bibitem{Lochmann}A.Lochmann: Diploma Thesis, Goettingen 2005, forthcoming

}

\end{thebibliography}
\end{document}